\documentclass[journal]{IEEEtran}
\usepackage{graphicx,amsmath,url}
\usepackage{microtype,lmodern}
\usepackage{mathtools,amssymb,gensymb}
\usepackage{tgtermes} 
\usepackage[T1]{fontenc}
\usepackage[utf8]{inputenc}
\usepackage{color,soul}
\usepackage[colorinlistoftodos]{todonotes}
\usepackage{acronym}
\usepackage{array}
\usepackage{booktabs}
\usepackage[compress]{cite}

\usepackage[normalem]{ulem}
\usepackage[switch]{lineno}
\usepackage{listings}
 
\usepackage{longtable}
\usepackage[colorlinks,breaklinks]{hyperref} 

\hypersetup{hypertexnames=true,
  linkcolor=blue,anchorcolor=black,citecolor=blue,urlcolor=blue
}
\graphicspath{{imglocal/}{img/}}
\acrodefplural{RAM}[RAMs]{Random Access Memories}
\acrodefplural{RRAM}[R-RAMs]{Resistive Random Access Memories}
\acrodefplural{STT-MRAM}[STT-MRAMs]{Spin-Transfer Torque Magnetic Random Access Memories}
\acrodef{ADC}[ADC]{Analog to Digital Converter}
\acrodef{ADEX}[AdExp-I\&F]{Adaptive-Exponential Integrate and Fire}
\acrodef{AER}[AER]{Address-Event Representation}
\acrodef{AEX}[AEX]{AER EXtension board}
\acrodef{AE}[AE]{Address-Event}
\acrodef{AFM}[AFM]{Atomic Force Microscope}
\acrodef{AGC}[AGC]{Automatic Gain Control}
\acrodef{AMDA}[AMDA]{AER Motherboard with D/A converters}
\acrodef{ANN}[ANN]{Attractor Neural Network}
\acrodef{API}[API]{Application Programming Interface}
\acrodef{ARM}[ARM]{Advanced RISC Machine}
\acrodef{ASIC}[ASIC]{Application Specific Integrated Circuit}
\acrodef{BCM}[BMC]{Bienenstock-Cooper-Munro}
\acrodef{BD}[BD]{Bundled Data}
\acrodef{BEOL}[BEOL]{Back-end of Line}
\acrodef{BG}[BG]{Bias Generator}
\acrodef{BMI}[BMI]{Brain-Machince Interface}
\acrodef{CAD}[CAD]{Computer Aided Design}
\acrodef{CAM}[CAM]{Content Addressable Memory}
\acrodef{CAVIAR}[CAVIAR]{Convolution AER Vision Architecture for Real-Time}
\acrodef{CFC}[CFC]{Current to Frequency Converter}
\acrodef{CCN}[CCN]{Cooperative and Competitive Network}
\acrodef{CHP}[CHP]{Communicating Hardware Processes}
\acrodef{CNN}[CCN]{Convolutional Neural Network}
\acrodef{CMIM}[CMIM]{Metal-insulator-metal Capacitor}
\acrodef{CMOL}[CMOL]{``Hybrid CMOS nanoelectronic circuits''}
\acrodef{CMOS}[CMOS]{Complementary Metal-Oxide-Semiconductor}
\acrodef{COTS}[COTS]{Commercial Off-The-Shelf}
\acrodef{CPG}[CPG]{Central Pattern Generator}
\acrodef{CPLD}[CPLD]{Complex Programmable Logic Device}
\acrodef{CPU}[CPU]{Central Processing Unit}
\acrodef{CV}[CV]{Coefficient of Variation}
\acrodef{DAC}[DAC]{Digital to Analog Converter}
\acrodef{DAS}[DAS]{Dynamic Auditory Sensor}
\acrodef{DAVIS}[DAVIS]{Dynamic and Active Pixel Vision Sensor}
\acrodef{DBN}[DBN]{Deep Belief Network}
\acrodef{DFA}[DFA]{Deterministic Finite Automaton}
\acrodef{DMA}[DMA]{Direct Memory Access}
\acrodef{DNF}[DNF]{Dynamic Neural Field}
\acrodef{DNN}[DNN]{Deep Neural Network}
\acrodef{DOF}[DOF]{Degrees of Freedom}
\acrodef{DPE}[DPE]{Dynamic Parameter Estimation}
\acrodef{DPI}[DPI]{Differential Pair Integrator}
\acrodef{DRAM}[DRAM]{Dynamic Random Access Memory}
\acrodef{DR}[DR]{Dual Rail}
\acrodef{DSP}[DSP]{Digital Signal Processor}
\acrodef{DVS}[DVS]{Dynamic Vision Sensor}
\acrodef{EBL}[EBL]{Electron Beam Lithography}
\acrodef{EDVAC}[EDVAC]{Electronic Discrete Variable Automatic Computer}
\acrodef{EIN}[EIN]{Excitatory-Inhibitory Network}
\acrodef{EM}[EM]{Expectation Maximization}
\acrodef{EPSC}[EPSC]{Excitatory Post-Synaptic Current}
\acrodef{EPSP}[EPSP]{Excitatory Post-Synaptic Potential}
\acrodef{FDSOI}[FD-SOI]{Fully-Depleted Silicon on Insulator}
\acrodef{FET}[FET]{Field-Effect Transistor}
\acrodef{FFT}[FFT]{Fast Fourier Transform}
\acrodef{FI}[F-I]{Frequency-Current}
\acrodef{FPGA}[FPGA]{Field Programmable Gate Array}
\acrodef{FSA}[FSA]{Finite State Automaton}
\acrodef{FSM}[FSM]{Finite State Machine}
\acrodef{GOPS}[GOPS]{Giga-Operations per Second}
\acrodef{GPU}[GPU]{Graphical Processing Unit}
\acrodef{GUI}[GUI]{Graphical User Interface}
\acrodef{HAL}[HAL]{Hardware Abstraction Layer}
\acrodef{HH}[H\&H]{Hodgkin \& Huxley}
\acrodef{HMM}[HMM]{Hidden Markov Model}
\acrodef{HRS}[HRS]{High-Resistive State}
\acrodef{HR}[HR]{Human Readable}
\acrodef{HSE}[HSE]{Handshaking Expansion}
\acrodef{HW}[HW]{Hardware}
\acrodef{ICT}[ICT]{Information and Communication Technology}
\acrodef{IC}[IC]{Integrated Circuit}
\acrodef{IF2DWTA}[IF2DWTA]{Integrate \& Fire 2--Dimensional WTA}
\acrodef{IFSLWTA}[IFSLWTA]{Integrate \& Fire Stop Learning WTA}
\acrodef{IF}[I\&F]{Integrate-and-Fire}
\acrodef{IMU}[IMU]{Inertial Measurement Unit}
\acrodef{INCF}[INCF]{International Neuroinformatics Coordinating Facility}
\acrodef{INI}[INI]{Institute of Neuroinformatics}
\acrodef{IO}[I/O]{Input/Output}
\acrodef{IoT}[IoT]{Internet of Things}
\acrodef{IPSC}[IPSC]{Inhibitory Post-Synaptic Current}
\acrodef{IPSP}[IPSP]{Inhibitory Post-Synaptic Potential}
\acrodef{IP}[IP]{Intellectual Property}
\acrodef{ISI}[ISI]{Inter-Spike Interval}
\acrodef{JFLAP}[JFLAP]{Java - Formal Languages and Automata Package}
\acrodef{LLC}[LLC]{Low Leakage Cell}
\acrodef{LFP}[LFP]{Local Field Potential}
\acrodef{LNA}[LNA]{Low-Noise Amplifier}
\acrodef{LPF}[LPF]{Low-Pass Filter}
\acrodef{LRS}[LRS]{Low-Resistive State}
\acrodef{LSM}[LSM]{Liquid State Machine}
\acrodef{LTD}[LTD]{Long Term Depression}
\acrodef{LTI}[LTI]{Linear Time-Invariant}
\acrodef{LTP}[LTP]{Long Term Potentiation}
\acrodef{LTU}[LTU]{Linear Threshold Unit}
\acrodef{LUT}[LUT]{Look-Up Table}
\acrodef{MCMC}[MCMC]{Markov-Chain Monte Carlo}
\acrodef{MEMS}[MEMS]{Micro Electro Mechanical System}
\acrodef{MIM}[MIM]{Metal Insulator Metal}
\acrodef{MOSCAP}[MOSCAP]{Metal Oxide Semiconductor Capacitor}
\acrodef{MOSFET}[MOSFET]{Metal Oxide Semiconductor Field-Effect Transistor}
\acrodef{MOS}[MOS]{Metal Oxide Semiconductor}
\acrodef{MRI}[MRI]{Magnetic Resonance Imaging}
\acrodef{NDFSM}[NDFSM]{Non-deterministic Finite State Machine} 
\acrodef{ND}[ND]{Noise-Driven}
\acrodef{NEF}[NEF]{Neural Engineering Framework}
\acrodef{NHML}[NHML]{Neuromorphic Hardware Mark-up Language}
\acrodef{NIL}[NIL]{Nano-Imprint Lithography}
\acrodef{NMDA}[NMDA]{N-Methyl-D-Aspartate}
\acrodef{NME}[NE]{Neuromorphic Engineering}
\acrodef{OTA}[OTA]{Operational Transconductance Amplifier}
\acrodef{PCB}[PCB]{Printed Circuit Board}
\acrodef{PFM}[PFM]{Pulse Frequency Modulation}
\acrodef{PR}[PR]{Production Rule}
\acrodef{PSC}[PSC]{Post-Synaptic Current}
\acrodef{PSTH}[PSTH]{Peri-Stimulus Time Histogram}
\acrodef{QDI}[QDI]{Quasi Delay Insensitive}
\acrodef{RAM}[RAM]{Random Access Memory}
\acrodef{RELU}[ReLU]{Rectified Linear Unit}
\acrodef{RMSE}[RMSE]{Root Mean Squared-Error}
\acrodef{RMS}[RMS]{Root Mean Squared}
\acrodef{RNN}[RNN]{Recurrent Neural Network}
\acrodef{ROLLS}[ROLLS]{Reconfigurable On-Line Learning Spiking}
\acrodef{RRAM}[RRAM]{Resistive Random Access Memory}
\acrodef{SAC}[SAC]{Selective Attention Chip}
\acrodef{SCX}[SCX]{Silicon CorteX}
\acrodef{SD}[SD]{Signal-Driven}
\acrodef{SEM}[SEM]{Spike-based Expectation Maximization}
\acrodef{SLAM}[SLAM]{Simultaneous Localization and Mapping}
\acrodef{SOC}[SOC]{System-On-Chip}
\acrodef{SOI}[SOI]{Silicon on Insulator}
\acrodef{SRAM}[SRAM]{Static Random Access Memory}
\acrodef{STDP}[STDP]{Spike-Timing Dependent Plasticity}
\acrodef{STD}[STD]{Short-Term Depression}
\acrodef{STP}[STP]{Short-Term Plasticity}
\acrodef{STT-MRAM}[STT-MRAM]{Spin-Transfer Torque Magnetic Random Access Memory}
\acrodef{STT}[STT]{Spin-Transfer Torque}
\acrodef{SW}[SW]{Software}
\acrodef{TFT}[TFT]{Thin Film Transistor}
\acrodef{USB}[USB]{Universal Serial Bus}
\acrodef{VHDL}[VHDL]{VHSIC Hardware Description Language}
\acrodef{VLSI}[VLSI]{Very Large Scale Integration}
\acrodef{VOR}[VOR]{Vestibulo-Ocular Reflex}
\acrodef{WTA}[WTA]{Winner-Take-All}
\acrodef{XML}[XML]{eXtensible Mark-up Language}
\acrodef{divmod3}[DIVMOD3]{divisibility of a number by 3}
\acrodef{hWTA}[hWTA]{Hard Winner-Take-All}
\acrodef{sWTA}[sWTA]{soft Winner-Take-All}

\begin{document}
%
\title{Analog circuits for mixed-signal neuromorphic computing architectures in 28\,nm FD-SOI technology}

\author{\IEEEauthorblockN{Ning Qiao\\}
\IEEEauthorblockA{Institute of Neuroinformatics\\University of Zurich and ETH Zurich\\Zurich, Switzerland\\
Email: qiaoning@ini.uzh.ch}
  \and
  \IEEEauthorblockN{Giacomo Indiveri}
\IEEEauthorblockA{Institute of Neuroinformatics\\University of Zurich and ETH Zurich\\Zurich, Switzerland\\
Email: giacomo@ini.uzh.ch}
}


\maketitle

\begin{abstract}
  Developing mixed-signal analog-digital neuromorphic circuits in advanced scaled processes poses significant design challenges. We present compact and energy efficient sub-threshold analog synapse and neuron circuits, optimized for a 28\,nm \acs{FDSOI} process, to implement massively parallel large-scale neuromorphic computing systems. We describe the techniques used for maximizing density with mixed-mode analog/digital synaptic weight configurations, and the methods adopted for minimizing the effect of channel leakage current, in order to implement efficient analog computation based on pA-nA small currents. We present circuit simulation results, based on a new chip that has been recently taped out, to demonstrate how the circuits can be useful for both low-frequency operation in systems that need to interact with the environment in real-time, and for high-frequency operation for fast data processing in different types of spiking neural network architectures.
\end{abstract}


\IEEEpeerreviewmaketitle

\section{Introduction}

As computing systems implemented in advanced \ac{VLSI} processes are facing more and more stringent requirements, mainly related power consumption, circuit designers and system engineers are starting to explore solutions that are alternative to the standard approach of the von Neumann computing paradigm. Neuromorphic computing represents one of these approaches, that proposes to use brain inspired neural network architectures for signal and data processing~\cite{Mead90,Indiveri_Horiuchi11}. One of the main features of neuromorphic computing architectures is their co-localization of memory and computing elements~\cite{Indiveri_Liu15}: the synapse elements in these neuromorphic architectures represent at the same time the site of memory (that store the synaptic weight value), and the site of computation (which in the simplest case is the multiplication of the input signal with the stored weight). From the computing architecture point of view, this has the large advantage of avoiding the von Neumann bottleneck problem~\cite{Backus78,Indiveri_Liu15}. Since memory transfer is typically the largest power consuming operation, this approach represents already a large step toward the development of ultra-low power computing systems. Another brain-inspired approach that is extremely helpful in reducing power consumption and increasing circuit density is that of representing signals using pulse-frequency modulation: if input/output signals are represented as pulses (spikes), the multiplication operation between input signals and synaptic weights reduces to a gating operation at the synapse level, that typically produces a weighted current at the arrival of the pre-synaptic spike that is integrated by the post-synaptic neuron. The higher the frequency of the input spikes, the larger the integrated value that the neuron sees. Furthermore, if many synapses receive input spikes in parallel, the weighted sum operation is implemented directly at the input node of the post-synaptic neuron by Kirchhoff's current law. Power consumption can be further reduced by implementing this spike- or event-based signal representation using asynchronous logic. In this case, the representation is denoted as \ac{AER}. Given these features, and given that this representation is also optimal for transmitting signals across long distances or chip boundaries, most of the recent state-of-the-art neuromorphic computing approaches use \ac{AER}~\cite{Benjamin_etal14,Furber_etal14,Qiao_etal15,Park_etal16,Merolla_etal14a}.
The last step that can be taken to further minimize power consumption is that of adopting a mixed-signal design approach, and using analog circuits that directly exploit the physics of the devices to implement the desired neural network computational primitives~\cite{Mead90}. As these primitives are mainly composed of exponential and logarithm functions, the best approach to follow is that of using sub-threshold analog circuits~\cite{Liu_etal02a}.

In this paper we present sub-threshold analog synapse and neuron circuits that have been designed to implement large-scale multi-neuron multi-core neuromorphic computing architectures using a 28\,nm \ac{FDSOI} process~\cite{Qiao_Indiveri16}. We show how it is possible to implement complex bio-physically realistic synaptic and neural dynamics using ultra-low power compact analog circuits in advanced scaled processes, by using mismatch-reducing and leakage-canceling techniques. Although the  circuits proposed can be configured to run at speeds that are much higher than biologically plausible ones, they have been optimized for reproducing time-constants that can be as long as tens or hundreds of milli-seconds. In this way these circuits can be embedded in massively parallel neural architectures optimally suited for processing live streaming data coming from natural sensory signals, such as auditory signals representing speech, visual signals representing gestures, or bio-signals measured from real neurons or muscles. Although the time constants of the individual computing elements are long, the asynchronous nature of the \ac{AER} protocol used to process incoming data ensures that latency and response time at the system level are extremely fast, i.e., ranging from hundreds of nano-seconds to milliseconds, depending on the complexity of the networks implemented. The circuits proposed represent a natural extension of similar circuits already fabricated, tested, and validated in less advanced processes~\cite{Qiao_etal15}, currently being used for implementing brain-machine interfaces~\cite{Corradi_Indiveri15,Boi_etal16}, deep network prototypes~\cite{Indiveri_etal15}, and autonomous driving robotic applications~\cite{Milde_etal17}.

\section{Sub-threshold analog/digital synapse \& neuron circuits}
\label{sec:silicon}

\begin{figure*}
  \centering
  \includegraphics[width=0.75\textwidth]{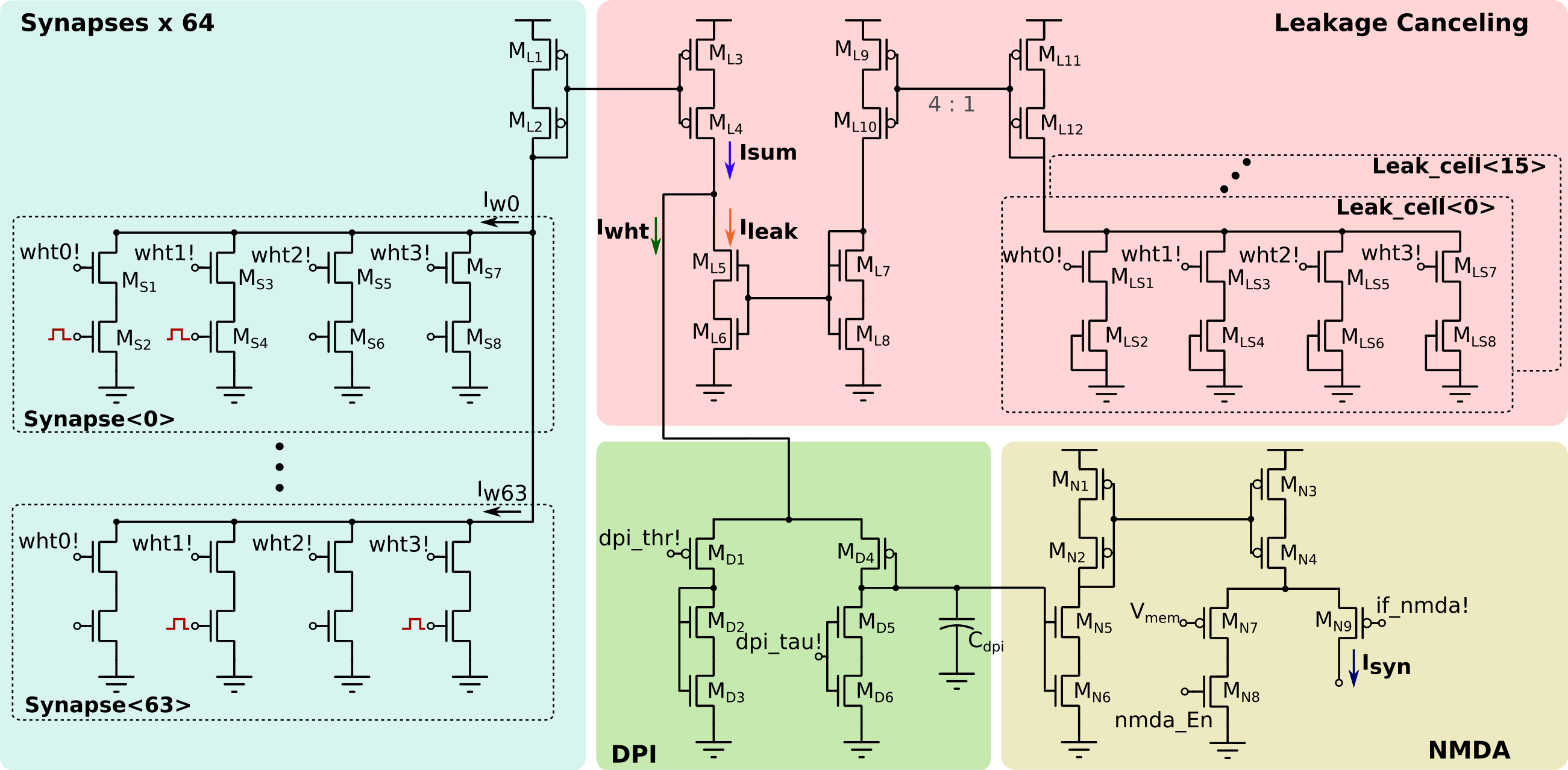}
  \caption{Schematic diagram of synapse and integrator circuits.}
  \label{fig:synapse}
\end{figure*}

Event-based synaptic circuits typically translate pre-synaptic voltage pulses into post-synaptic currents and source them into the target neuron circuit with a gain that corresponds to the synaptic weight. In Fig.~\ref{fig:synapse} we show the schematic diagram of a circuit that comprises 64 programmable synapse blocks with common leakage compensation and temporal dynamics blocks. 

Each synapse has 4 current branches with shared analog bias settings \textsf{wht\,x!}, that can be programmed with a 10-bit temperature compensated bias generator~\cite{Yang_etal12}. In the circuit diagrams, all signals ending with a "!" represent programmable bias settings. Upon the arrival of a pre-synaptic spike, the input \ac{AER} event is decoded to select the activation of one or more of the synapse branches, therefore allowing to set as synaptic weight one of 16 possible analog currents.

In advanced scaled processes, such as the one used in this work, the off-channel leakage current of each branch is of the order of pico-Amperes or more. Considering that there are 64 synapse blocks with a total of 256 current branches, the total leakage current produced is non-negligible. The ``Leakage Canceling'' block of Fig.~\ref{fig:synapse} attempts to produce the same leakage current, with a 4-to-1 copy of 16 leak cells with 4 branches each that are biased with the same settings used for the synapse blocks. All dark currents from the leak cells branches are summed and copied by $\mathsf{M_{L5}-M_{L12}}$ of Fig.~\ref{fig:synapse} to produce the $\mathsf{I_{leak}}$ current. The ``Leakage Canceling'' block subtracts this current from the total synaptic current $\mathsf{I_{sum}}$, to produce a resulting output current $\mathsf{I_{wht}}$ which represents the compensated weighted contribution of all 64 afferent synapses. If one assumes that all synapses share the same dynamics, it is possible to use the superposition principle and model the temporal dynamics of all synapses using one single low pass filter. This is indeed the case for the circuits of Fig.~\ref{fig:synapse}, that use the \ac{DPI} block on the bottom right of the figure to implement a current-mode low-pass filter that exhibits synaptic dynamics~\cite{Bartolozzi_Indiveri07a}. The time constant of this circuit is directly proportional to its capacitance, and inversely proportional to the current through $\mathsf{M_{D5}-M_{D6}}$. In order to obtain large time constants, while keeping the size of the capacitors to a minimum, it is necessary to generate bias currents that can be as small as pico-Amperes. To achieve this goal in this process, we had to resort to the use of the "pseudo-cascode" split-transistor sub-threshold technique~\cite{Saxena_Baker08} (e.g., see also $\mathsf{M_{L1}-M_{L4}}$ in Fig.~\ref{fig:synapse}).  The diode connected transistors $\mathsf{M_{D2}-M_{D3}}$ are added to reduce the $\mathsf{V_{DS}}$ of $\mathsf{M_{D1}}$ so as to reduce its early effect, and to improve the circuit's linear performance.
The \emph{NMDA} block of Fig.~\ref{fig:synapse} ($\mathsf{M_{N1}-M_{N3}}$) models the voltage-gating mechanisms of \ac{NMDA} synapses, which can be useful for coincidence detection and precise spike-timing signal processing strategies in spike-based neuromorphic computing applications. The area of the synapse block layout is 3\,$\mu m^{2}$, and the active area of the \ac{DPI} circuit is 12.5\,$\mu m^{2}$. The \ac{DPI} capacitor, implemented using  a MIMCAP structure covering the neighboring circuits, measures 1\,pF.

\begin{figure*}
  \centering
  \includegraphics[width=0.75\textwidth]{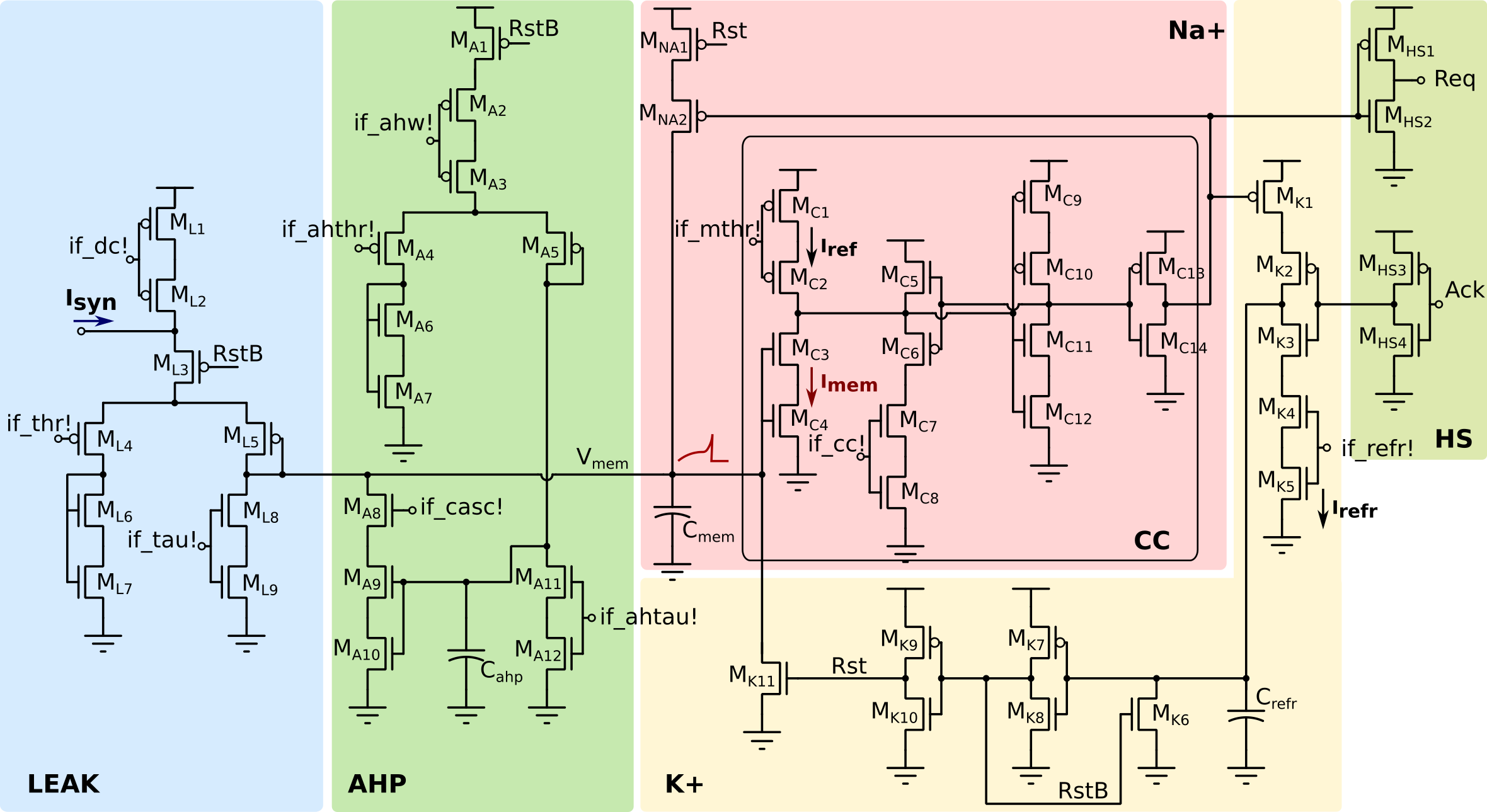}
  \caption{Schematic diagram of an analog \ac{IF} neuron.}
  \label{fig:neuron}
\end{figure*}

The circuit that implements the silicon neuron functionality is shown in Fig.~\ref{fig:neuron}. This is a current-mode circuit composed of multiple ``compartments'' or blocks. The \emph{LEAK} block 
models the neuron's passive leak conductance, producing exponential sub-threshold dynamics in response to constant input currents. The \emph{AHP} block 
models the generation of the after hyper-polarizing current in real neurons, responsible for their spike-frequency adaptation behavior. The Na+ 
and K+ blocks 
model the effect of Sodium and Potassium channels, responsible for generating action-potentials (spikes) in real neurons. The \emph{HS} block 
implements handshaking with following encoder block for encoding spike events following \ac{AER} protocol. 
We used an optimized Traff's current comparator (see \emph{CC} box in the Na+ block) to make an accurate comparison between the neuron $\mathsf{I_{mem}}$ current and a programmable  $\mathsf{I_{ref}}$ threshold current, which sets the neuron's spiking threshold. Current limitation transistors ($\mathsf{M_{C7},M_{C8}}$) are included to reduce static power consumption. We used the same split-transistor sub-threshold technique used in the synapse circuits for enhanced current-mirror operation and for precise control of  small currents. We added also several reset transistors, such as $\mathsf{M_{L3}}$ and $\mathsf{M_{NA1}}$ to further reduce power consumption during the spike reset phase. The active area of the neuron is 20\,$\mu m^{2}$; the neuron capacitance is also implemented using a MIMCAP structure and measures approximately 1.5\,pF. Fig.~\ref{fig:isynimem} shows the expected response of the neuron circuit to a constant input current. By tuning the biases that control the neuron's integration time constant, firing threshold, refractory period and spike-frequency adaptation dynamics, the proposed circuit can reproduce a wide range of spiking behaviors~\cite{Qiao_etal15}.

\begin{figure}
  \centering
  \includegraphics[width=0.35\textwidth]{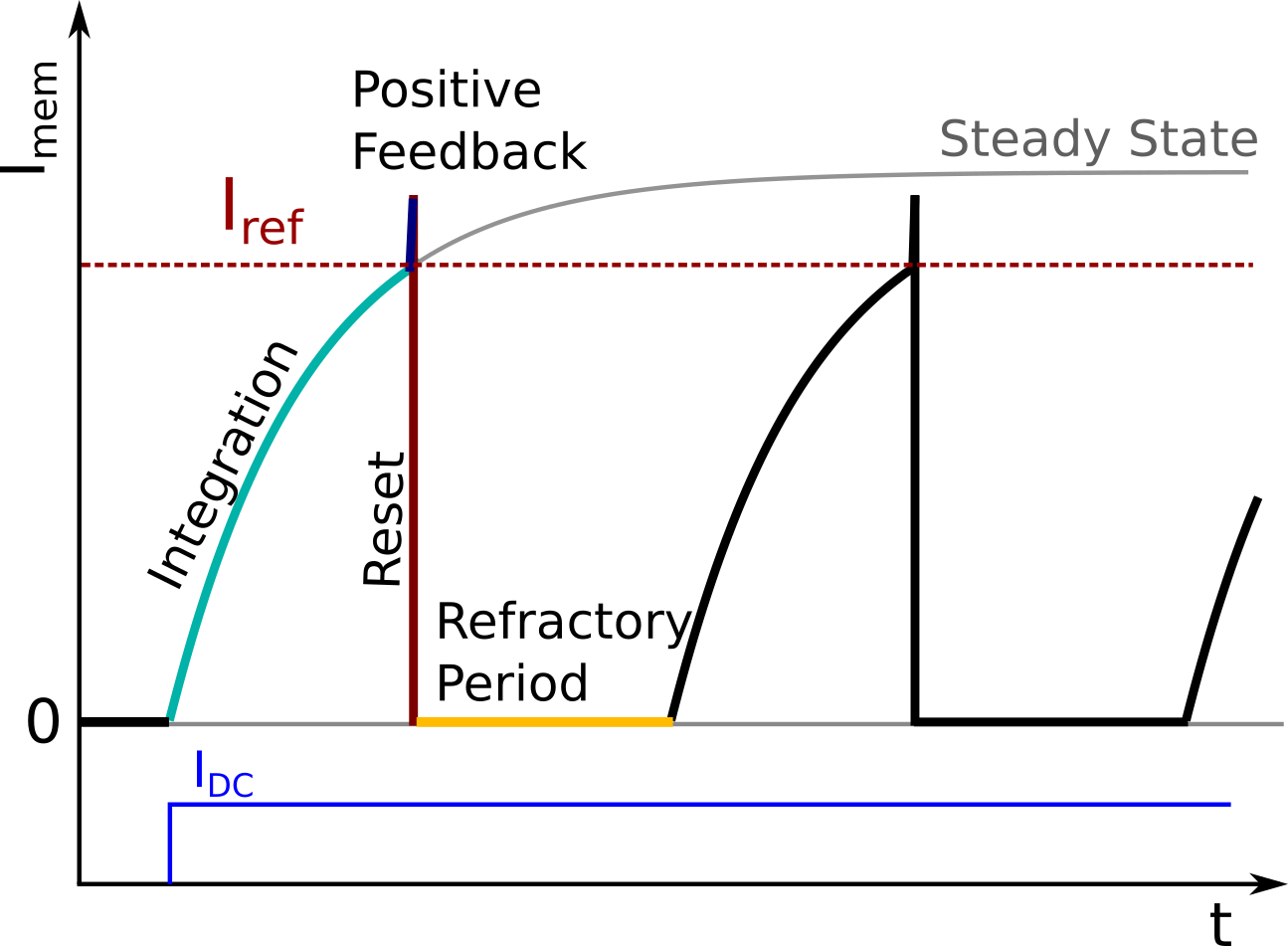}
  \caption{Membrane current trace over time.}
  \label{fig:trace}
\end{figure}



\section{Simulation results}
\label{sec:results}

\begin{figure}
  \centering
  \includegraphics[width=0.4\textwidth]{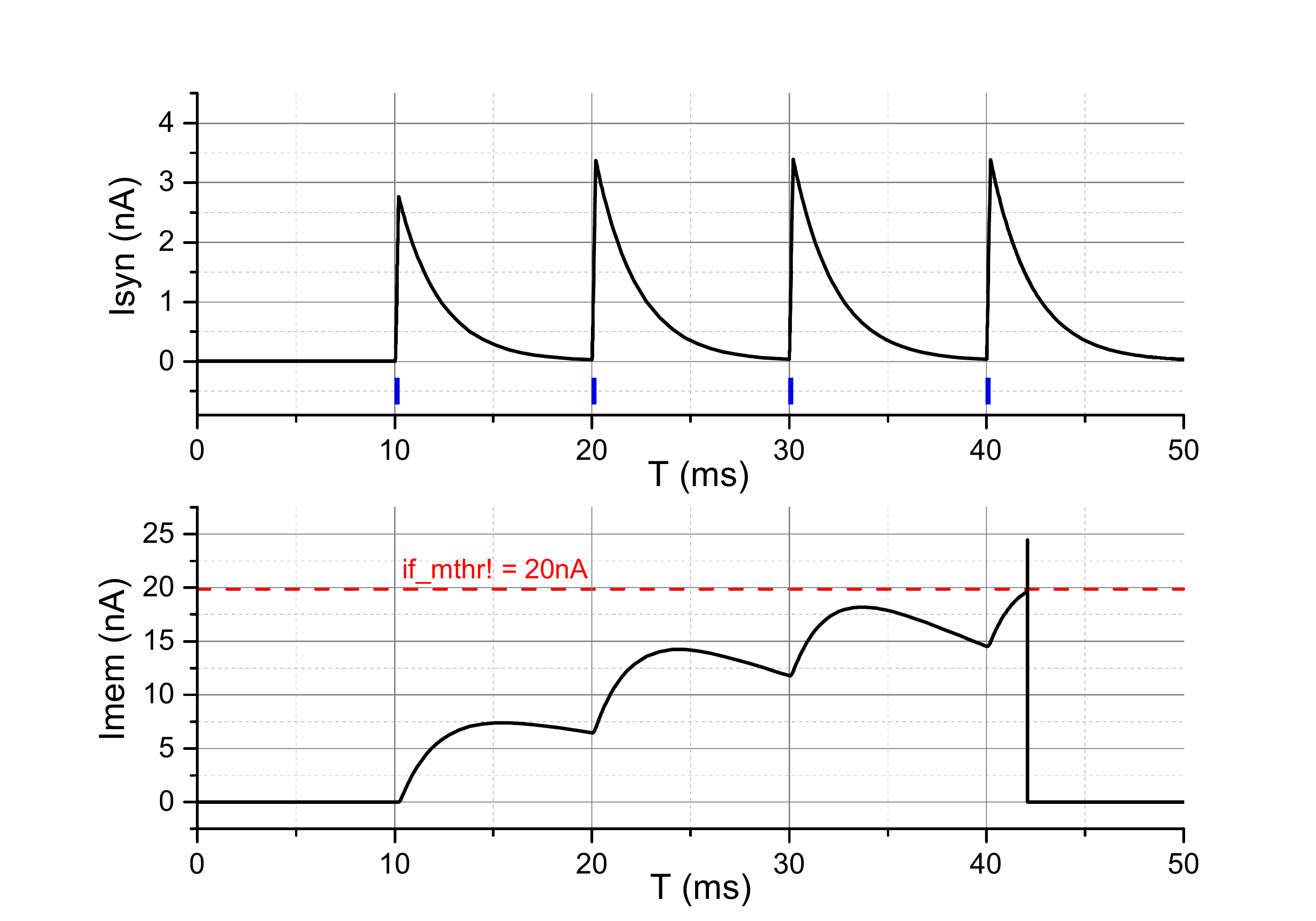}
  \caption{Synapse and neuron response to a 100\,Hz spike train.}
  \label{fig:isynimem}
\end{figure}

In Fig.~\ref{fig:isynimem} we show circuit simulation results of both synaptic and neuron currents, while they are being stimulated by a 100\,Hz input spike train with pulse width of 200\,us and pulse amplitude of 10\,nA. The spiking threshold reference current was set to 
20\,nA. As shown, both synaptic and neuron output signals exhibit biologically plausible time temporal dynamics with time constants of the order of milli-seconds.

\begin{figure}
  \centering
  \includegraphics[width=0.375\textwidth]{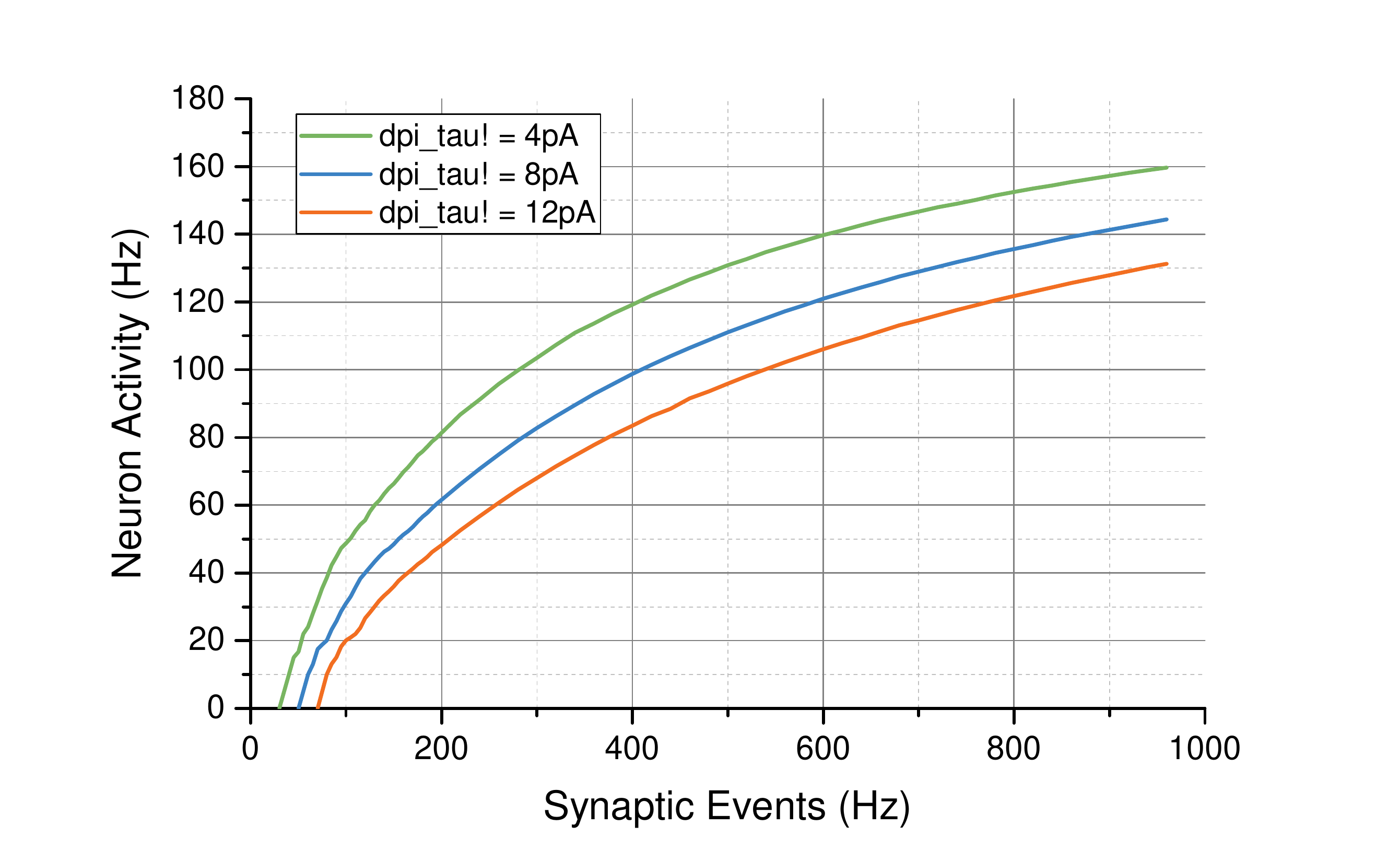}
  \caption{Combined synapse-neuron transfer function with refractory period of 5ms for synaptic input firing rate from 0-1k\,Hz. }
  \label{fig:sigmoid}
\end{figure}

\begin{figure}
  \centering
  \includegraphics[width=0.375\textwidth]{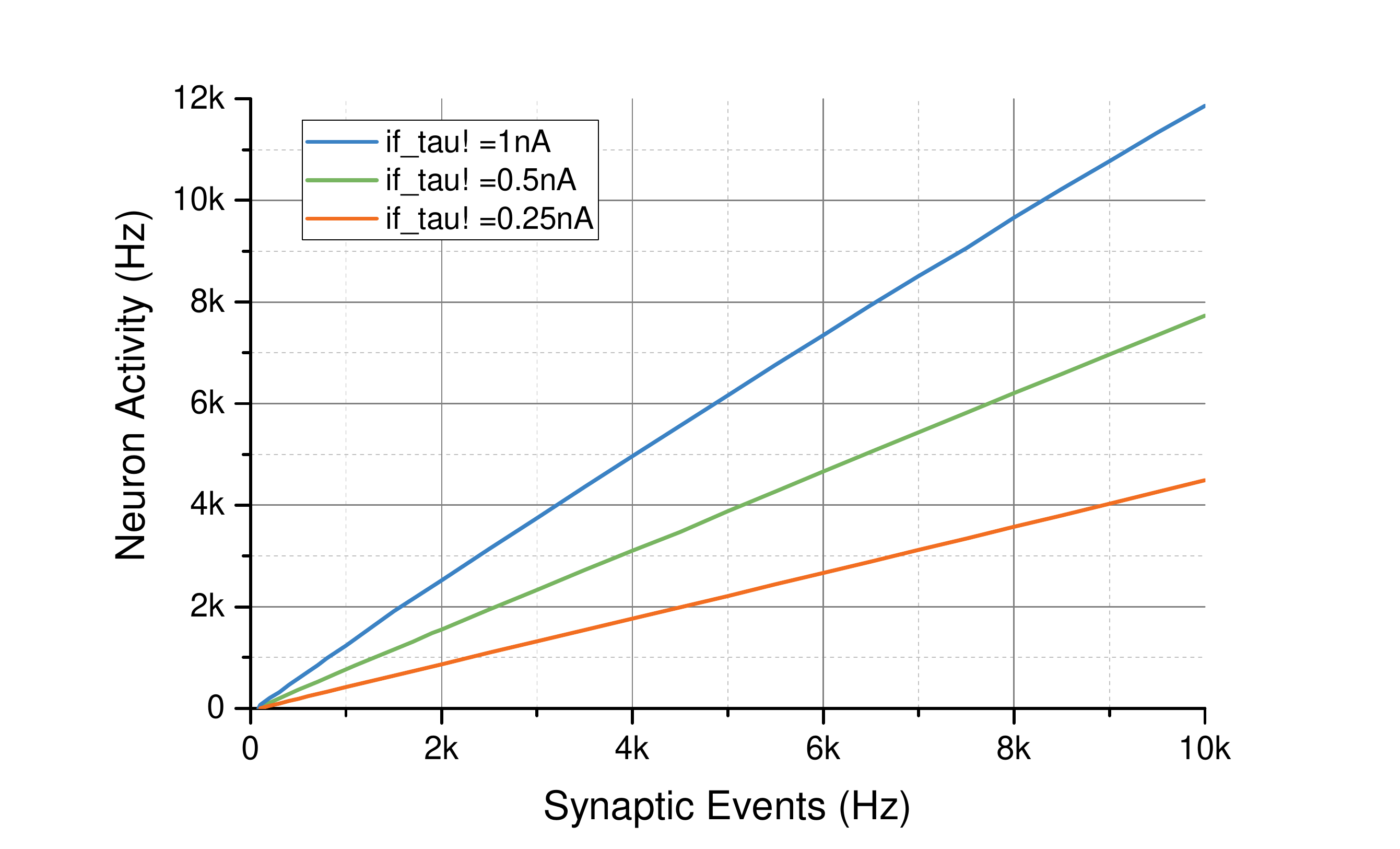}
  \caption{Combined synapse-neuron transfer function with short refractory period.}
  \label{fig:relu}
\end{figure}

Fig.~\ref{fig:sigmoid} shows the combined synapse-neuron transfer function, consisting of the neuron output firing rate as a function of the synapse input firing rate, for three different synaptic efficacy levels (which are inversely proportional to the $\mathsf{dpi\_tau}$! bias setting~\cite{Qiao_etal15}). In these simulations the synaptic pulse width was set to 1\,ms, the pulse amplitude to 10\,nA, the neuron spiking threshold reference to 20\,nA, and its refractory period to 5\,ms. Setting a refractory period to long intervals forces the neuron circuits to saturate at low frequencies, therefore reproducing the behavior of real neurons and limiting the bandwidth requirement for spiking neural networks. By changing the bias settings that affect the neuron refractory period, it is possible to configure the circuit to operate in a linear manner over a much larger range of output frequencies. Fig.~\ref{fig:relu} shows the same transfer function with bias settings tuned to reproduce the function of \ac{RELU} units, typically used in deep-networks, over a wide range of fast input/output frequencies, thus making these circuits also suitable for high-speed spiking deep network models. The data of Fig.~\ref{fig:relu} was obtained by setting the width of the input synaptic pulses to 50\,us, their amplitude to 10\,nA, and the refractory period to micro-seconds. The slope of the transfer function can be modulated by changing the gain of the synapse/neuron \ac{DPI} circuits, via the corresponding $\mathsf{\_thr}$! and $\mathsf{\_tau}$! bias settings of Fig.~\ref{fig:synapse} and of Fig.~\ref{fig:neuron}.

\section{Conclusion}

We proposed novel compact analog/digital neuromorphic circuits optimized for minimizing leakage currents and producing long time constants in a 28\,nm \ac{FDSOI} process, that can enable neuromorphic architectures to process natural sensory signals in real-time. This allows the design of massively parallel ultra-low power mixed-signal neuromorphic computing architectures that would not be affected by the von Neumann bottleneck, as they would not need to time-multiplex shared neuron/synapse circuits and transfer their state information to separate memory blocks. We showed how the circuits proposed reproduce the synapse and neural dynamics expected from theory and can be used to reproduce both biologically realistic dynamics, or fast \ac{RELU} transfer functions. They are fully compatible with spike-based learning algorithm/circuits~\cite{Qiao_etal15} and can be readily integrated in the next generation of large multi-neuron, multi-core neuromorphic architectures.

\section*{Acknowledgment}
This work is supported by the EU ICT grant ``NeuRAM$^3$'' (687299).

\bibliographystyle{IEEEtran}

\begin{thebibliography}{10}
\providecommand{\url}[1]{#1}
\csname url@samestyle\endcsname
\providecommand{\newblock}{\relax}
\providecommand{\bibinfo}[2]{#2}
\providecommand{\BIBentrySTDinterwordspacing}{\spaceskip=0pt\relax}
\providecommand{\BIBentryALTinterwordstretchfactor}{4}
\providecommand{\BIBentryALTinterwordspacing}{\spaceskip=\fontdimen2\font plus
\BIBentryALTinterwordstretchfactor\fontdimen3\font minus
  \fontdimen4\font\relax}
\providecommand{\BIBforeignlanguage}[2]{{%
\expandafter\ifx\csname l@#1\endcsname\relax
\typeout{** WARNING: IEEEtran.bst: No hyphenation pattern has been}%
\typeout{** loaded for the language `#1'. Using the pattern for}%
\typeout{** the default language instead.}%
\else
\language=\csname l@#1\endcsname
\fi
#2}}
\providecommand{\BIBdecl}{\relax}
\BIBdecl

\bibitem{Mead90}
C.~Mead, ``Neuromorphic electronic systems,'' \emph{Proceedings of the {IEEE}},
  vol.~78, no.~10, pp. 1629--36, 1990.

\bibitem{Indiveri_Horiuchi11}
\BIBentryALTinterwordspacing
G.~Indiveri and T.~Horiuchi, ``Frontiers in neuromorphic engineering,''
  \emph{Frontiers in Neuroscience}, vol.~5, no. 118, pp. 1--2, 2011. [Online].
  Available:
  \url{http://www.frontiersin.org/neuromorphic_engineering/10.3389/fnins.2011.00118/full}
\BIBentrySTDinterwordspacing

\bibitem{Indiveri_Liu15}
\BIBentryALTinterwordspacing
G.~Indiveri and S.-C. Liu, ``Memory and information processing in neuromorphic
  systems,'' \emph{Proceedings of the {IEEE}}, vol. 103, no.~8, pp. 1379--1397,
  2015. [Online]. Available:
  \url{http://ncs.ethz.ch/pubs/pdf/Indiveri_Liu15.pdf}
\BIBentrySTDinterwordspacing

\bibitem{Backus78}
\BIBentryALTinterwordspacing
J.~Backus, ``Can programming be liberated from the von neumann style?: a
  functional style and its algebra of programs,'' \emph{Communications of the
  ACM}, vol.~21, no.~8, pp. 613--641, 1978. [Online]. Available:
  \url{http://doi.acm.org/10.1145/359576.359579}
\BIBentrySTDinterwordspacing

\bibitem{Benjamin_etal14}
B.~V. Benjamin, P.~Gao, E.~McQuinn, S.~Choudhary, A.~R. Chandrasekaran,
  J.~Bussat, R.~Alvarez-Icaza, J.~Arthur, P.~Merolla, and K.~Boahen,
  ``Neurogrid: A mixed-analog-digital multichip system for large-scale neural
  simulations,'' \emph{Proceedings of the {IEEE}}, vol. 102, no.~5, pp.
  699--716, 2014.

\bibitem{Furber_etal14}
S.~Furber, F.~Galluppi, S.~Temple, and L.~Plana, ``The {SpiNNaker} project,''
  \emph{Proceedings of the IEEE}, vol. 102, no.~5, pp. 652--665, May 2014.

\bibitem{Qiao_etal15}
\BIBentryALTinterwordspacing
N.~Qiao, H.~Mostafa, F.~Corradi, M.~Osswald, F.~Stefanini, D.~Sumislawska, and
  G.~Indiveri, ``A re-configurable on-line learning spiking neuromorphic
  processor comprising 256 neurons and 128k synapses,'' \emph{Frontiers in
  Neuroscience}, vol.~9, no. 141, 2015. [Online]. Available:
  \url{http://www.frontiersin.org/neuromorphic_engineering/10.3389/fnins.2015.00141/abstract}
\BIBentrySTDinterwordspacing

\bibitem{Park_etal16}
J.~Park, T.~Yu, S.~Joshi, C.~Maier, and G.~Cauwenberghs, ``Hierarchical address
  event routing for reconfigurable large-scale neuromorphic systems,''
  \emph{{IEEE} Transactions on Neural Networks and Learning Systems}, pp.
  1--15, 2016.

\bibitem{Merolla_etal14a}
\BIBentryALTinterwordspacing
P.~A. Merolla, J.~V. Arthur, R.~Alvarez-Icaza, A.~S. Cassidy, J.~Sawada,
  F.~Akopyan, B.~L. Jackson, N.~Imam, C.~Guo, Y.~Nakamura, B.~Brezzo, I.~Vo,
  S.~K. Esser, R.~Appuswamy, B.~Taba, A.~Amir, M.~D. Flickner, W.~P. Risk,
  R.~Manohar, and D.~S. Modha, ``A million spiking-neuron integrated circuit
  with a scalable communication network and interface,'' \emph{Science}, vol.
  345, no. 6197, pp. 668--673, Aug 2014. [Online]. Available:
  \url{http://www.sciencemag.org/content/345/6197/668}
\BIBentrySTDinterwordspacing

\bibitem{Liu_etal02a}
\BIBentryALTinterwordspacing
S.-C. Liu, J.~Kramer, G.~Indiveri, T.~Delbruck, and R.~Douglas, \emph{Analog
  {VLSI}:Circuits and Principles}.\hskip 1em plus 0.5em minus 0.4em\relax MIT
  Press, 2002. [Online]. Available:
  \url{http://ncs.ethz.ch/pubs/pdf/Liu_etal02b.pdf}
\BIBentrySTDinterwordspacing

\bibitem{Qiao_Indiveri16}
\BIBentryALTinterwordspacing
N.~Qiao and G.~Indiveri, ``Scaling mixed-signal neuromorphic processors to 28nm
  fd-soi technologies,'' in \emph{Biomedical Circuits and Systems Conference,
  ({BioCAS}), 2016}.\hskip 1em plus 0.5em minus 0.4em\relax IEEE, 2016, pp.
  552--555. [Online]. Available:
  \url{http://ncs.ethz.ch/pubs/pdf/QiaoIndiveri16.pdf}
\BIBentrySTDinterwordspacing

\bibitem{Corradi_Indiveri15}
\BIBentryALTinterwordspacing
F.~Corradi and G.~Indiveri, ``A neuromorphic event-based neural recording
  system for smart brain-machine-interfaces,'' \emph{Biomedical Circuits and
  Systems, {IEEE} Transactions on}, vol.~9, no.~5, pp. 699--709, 2015.
  [Online]. Available: \url{http://ncs.ethz.ch/pubs/pdf/Corradi_Indiveri15.pdf}
\BIBentrySTDinterwordspacing

\bibitem{Boi_etal16}
\BIBentryALTinterwordspacing
F.~Boi, T.~Moraitis, V.~De~Feo, F.~Diotalevi, C.~Bartolozzi, G.~Indiveri, and
  A.~Vato, ``A bidirectional brain-machine interface featuring a neuromorphic
  hardware decoder,'' \emph{Frontiers in Neuroscience}, vol.~10, p. 563, 2016.
  [Online]. Available:
  \url{http://journal.frontiersin.org/article/10.3389/fnins.2016.00563}
\BIBentrySTDinterwordspacing

\bibitem{Indiveri_etal15}
\BIBentryALTinterwordspacing
G.~Indiveri, F.~Corradi, and N.~Qiao, ``Neuromorphic architectures for spiking
  deep neural networks,'' in \emph{Electron Devices Meeting {(IEDM)}, 2015
  {IEEE} International}.\hskip 1em plus 0.5em minus 0.4em\relax IEEE, Dec.
  2015, pp. 4.2.1--4.2.14. [Online]. Available:
  \url{http://ncs.ethz.ch/pubs/pdf/Indiveri_etal15.pdf}
\BIBentrySTDinterwordspacing

\bibitem{Milde_etal17}
M.~B. Milde, H.~Blum, A.~Dietm\"uller, H.~Blum, D.~Sumislawska, J.~Conradt,
  G.~Indiveri, and Y.~Sandamirskaya, ``Obstacle avoidance and target
  acquisition for robot navigation using a mixed signal analog/digital
  neuromorphic processing system,'' \emph{Frontiers in Neuroscience}, 2017.

\bibitem{Yang_etal12}
M.~Yang, S.-C. Liu, C.~Li, and T.~Delbruck, ``Addressable current reference
  array with 170db dynamic range,'' in \emph{Circuits and Systems ({ISCAS}),
  2012 {IEEE} International Symposium on}.\hskip 1em plus 0.5em minus
  0.4em\relax IEEE, 2012, pp. 3110--3113.

\bibitem{Bartolozzi_Indiveri07a}
\BIBentryALTinterwordspacing
C.~Bartolozzi and G.~Indiveri, ``Synaptic dynamics in analog {VLSI},''
  \emph{Neural Computation}, vol.~19, no.~10, pp. 2581--2603, Oct 2007.
  [Online]. Available:
  \url{http://ncs.ethz.ch/pubs/pdf/Bartolozzi_Indiveri07.pdf}
\BIBentrySTDinterwordspacing

\bibitem{Saxena_Baker08}
V.~Saxena and R.~J. Baker, ``Compensation of {CMOS} op-amps using split-length
  transistors,'' in \emph{Circuits and Systems ({MWSCAS}), 2008 {IEEE} 51st
  International Midwest Symposium on}.\hskip 1em plus 0.5em minus 0.4em\relax
  IEEE, 2008, pp. 109--112.

\end{thebibliography}

\end{document}